# Spin-polarization anisotropy included by mechanical bending in tungsten diselenide nanoribbons and tunable excitonic states

Hong Tang,*† Santosh Neupane,† Li Yin, Jason M. Breslin, and Adrienn Ruzsinszky

Department of Physics, Temple University, Philadelphia, PA 19122

**ABSTRACT** A WSe$_2$ monolayer shows many interesting properties due to its spin-orbit coupling induced spin splitting in bands around the Fermi level and the spin-valley configuration. The orientation of the spin polarization in the relevant bands is crucial for the nature of exciton states and the optical valley selectivity. In this work, we studied the WSe$_2$ nanoribbons under different mechanical bending curvatures and electron/hole doping with density functional theory and their optical absorption and excitonic states with many-body perturbation GW and BSE (Bethe-Salpeter equation) methods. We found that the WSe$_2$ nanoribbons can exhibit an enhanced SOC effect and a spatially varying spin polarization in bands around the Fermi level under bending. The spin-polarization can show an anisotropy (or asymmetry) in those nearly degenerate bands, leading to a controllable magnetism via bending and electron/hole doping to the nanoribbons, suggesting a potential application in compact and controllable magnetic nanodevices and spintronics. The optical absorption spectrum of the nanoribbon presents a large tunability with bending within the near infrared region of about 0.4 to 1.5 eV, showing an enhanced absorption at a large bending condition. The exciton states generally show mixed or various spin configuration in the electron and hole pairs that are controlled by bending, potentially useful for applications in spin-based quantum information processes.

## Introduction

Two-dimensional (2D) layered materials are an enduringly exciting area of research since the discovery of atomically thin graphene.[1-6] Recently, several experimental breakthroughs were achieved in 2D layered material systems,[7-16] especially in magnetic chromium halides[7-11] and versatile transition metal dichalcogenide (TMD) systems.[12-17] The controllable magneto-optical Faraday and Kerr effects,[7] spontaneous circularly polarized photoluminescence,[8] large linear magnetoelectric (ME) effect,[9,10] and gate-voltage controlled electron or hole doping for the tunability of magnetic ordering[11] were demonstrated in CrI$_3$ layered systems, suggesting its unique applicability for electrically or magnetically controlled magnetic nanodevices. Zhang et al.[12] showed that an in-plane magnetic field can brighten the spin-forbidden dark excitons in monolayer WSe$_2$ and precisely measured the energy levels of both the neutral and charged dark excitons in experiments. The largely increased emission and valley lifetimes of the magnetically brightened dark excitons permit storing and manipulating photoexcited valley and spin degrees of freedom for quantum information processes. Furthermore, Zhou et al.[17] showed the dark exciton with out-of-plane transition dipole moment in monolayer WSe$_2$ can emit energy to the surface plasmon polariton in the underlying metal substrate via near-field coupling, opening an avenue for probing and manipulating exciton dynamics in TMD monolayers.

---

\* Email: hongtang@temple.edu
† Authors having equal contributions to the work.

Nanoribbons are interesting, reduced forms of 2D layered materials.[18-26] Compared with their 2D counterparts, they have more spatial confinement effects and rich edge states, making them rife of new functionalities and attractive for applications in compact nano electronics, magnetic, optoelectronic, and band-topology based electronic and quantum information devices.[18,19] $MoS_2$ zigzag nanoribbons were calculated as metallic and with ferromagnetic edges.[20] The carbon atoms in each edge of graphene zigzag nanoribbons (GZNRs) are ferromagnetically coupled, while the inter-edge coupling changes from antiferromagnetic in narrow semiconductive GZNRs to ferromagnetic in wide metallic GZNRs.[19] Tang et al.[21] calculated that the direction of magnetization in $CrI_3$ nanoribbons can be stably aligned both out-of-plane and parallel to ribbon axis, increasing their operating controllability. Also, the rich exciton states were shown in SiC,[22] $MoS_2$,[23] black phosphorene,[24] and $CrI_3$ nanoribbons,[21] and the exciton states can be tuned by bending or strains in the nanoribbons.

It was shown[27] that in the 2H phase of $WSe_2$ monolayers, the spin-orbit coupling (SOC) effect induces a strong spin-splitting (~0.5 eV) around the valence band maximum (around the K point in the Brillouin zone) and a much smaller spin-splitting (~100 meV) around the conduction band minimum (also around the K point), while around the Γ point, the bands are spin degenerate. This results in opposite spin configurations in the two valleys around K and K′, crucial for the formation of the low energy dark exciton, the realization of the in-plane magnetic field induced dark exciton brightening and control, and the opposite circularly polarized emissions from the dark excitons.[12] TMD nanoribbons are cut from monolayers, where both the edge states and valence band continuum states are mainly derived from transition metal atoms, which usually exhibit a strong SOC effect. Bending can induce large non-uniform strains in nanoribbons, hence an enhanced internal electric field within nanoribbons. This may increase the SOC effect. In this work, we investigated the SOC effect and the spin polarization of bands around the Fermi level with density functional theory (DFT) for $WSe_2$ nanoribbons under different bending curvatures and doping, their optical absorption and excitonic states with many-body perturbation GW and BSE (Bethe-Salpeter equation) methods. We have found that the $WSe_2$ nanoribbons with bending can produce enhanced SOC effects and spatially varied spin polarization in bands around the Fermi level. The spin-polarization shows an anisotropy in those nearly degenerate bands, leading to a controllable magnetism with bending and having an important implication to the spin configuration of exciton states. Furthermore, electron or hole doping to nanoribbons can make the systems potentially applicable in compact and controllable magnetic nanodevices and spintronics. The optical absorption spectrum of the nanoribbon presents a large tunability with bending within the energy region of about 0.4 to 1.5 eV. The exciton states generally show mixed or various spin configuration in the electron and hole pairs, potentially beneficial for applications in spin-based quantum information processes.

**Results and discussion**

1. **Band gaps and enhanced SOC with bending**

The nanoribbon with armchair edges is cut from a $WSe_2$ monolayer of 2H phase in which tungsten (W) atoms relate to chalcogen selenium (Se) atoms in a trigonal prism, and the edge atoms are passivated with hydrogen (H) atoms, as shown in Figure 1a-c. The periodic (or ribbon length) direction of the built nanoribbons is aligned along the z axis, with their width directions along the x axis and the vacuum layer along the y axis. The supercell vectors a, b, and c are aligned with the axes x, y, and z, respectively. We use

A$n$WSe$_2$ to denote the WSe$_2$ nanoribbons, where $n$ represents the number of W atoms in the supercell and $n$=13 means 13 W atoms in the super cell, as shown in Figure 1a. The relaxed structures of A13WSe$_2$ under different bending curvature radii are shown in Figure 1d, and those for A12WSe$_2$ and A7WSe$_2$ are shown in Figures S1 and S2 of the supplemental information (SI). The calculated band structures of A13WSe$_2$ under different bending curvatures are shown in Figure 1e-n. The edge band gap (EG) is defined as the difference between the conduction band minimum (CBM) and the valence band maximum (VBM), while the non-edge band gap (NEG) is defined as the one between the bottom of the conduction band continuum (CBC) and the top of the valence band continuum (VBC), see Figure 1f. With large bending curvatures (R < 9 Å), the valence bands just below the Fermi level merge into VBC and in these cases, EG is the difference between CBM and VBC, as shown in Figure 1m. As can be seen, the flat A13WSe$_2$ has a direct NEG at Γ and a slightly indirect EG between Γ and Z, while under all bending curvature radii, both NEG and EG at Γ are direct. A12WSe$_2$ and A7WSe$_2$ also show a changing trend of direct or indirect NEG and EG with bending, as shown in SI Figure S15 for A12WSe$_2$ and SI Figure S22 for A7WSe$_2$. The changing trends of NEG and EG with bending curvatures $\kappa$ ($\kappa = 1/R$) for A12WSe$_2$ and A13WSe$_2$ reflect the feature. As shown in Figures 1o and 1p, A13WSe$_2$ (A12WSe$_2$) has an almost unchanging or slowly decreasing NEG up to $\kappa = 0.10/$Å ($\kappa = 0.11/$Å), then quickly decreasing NEG for $\kappa > 0.10/$Å ($\kappa > 0.11/$Å). The EG gap of A13WSe$_2$ shows a general decrease with $\kappa$, while within $0.071 /$Å $< \kappa < 0.125/$Å (or 8 Å < R < 14 Å) it shows a relatively flat or a slight hump feature, in line with the EG changing trend with $\kappa$ of the armchair MoS$_2$ nanoribbons in a previous study [23]. Due to the indirect EG of A12WSe$_2$ within $0.077 /$Å $< \kappa < 0.10/$Å (or 10 Å < R < 13 Å), the EG gap of A12WSe$_2$ shows a relatively large value and forms a peak within $0.071 /$Å $< \kappa < 0.11/$Å. The EG and NEG gaps of A7WSe$_2$ show a less varying feature with $\kappa$, as seen in Figures 1o and 1p.

The density of states (DOS) analysis for A13WSe$_2$ is shown in SI Figure S4, and those for A12WSe$_2$ and A7WSe$_2$ are in SI Figures S17 and S24. For A13WSe$_2$ and A12WSe$_2$, the total six W atoms located outmost and near to the edges, three W atoms for each edge, are counted as the edge W atoms, while for A7WSe$_2$, the total four outmost W atoms (two for each edge) are counted as the edge W atoms, see Figure 1a-c for the nanoribbon structures. For A13WSe$_2$ and A12WSe$_2$, for the flat nanoribbon and for low bending, the edge W atoms dominate the DOS around the Fermi level, see SI Figures S4 and S17. Those bands corresponding to those DOS (about 0 to 0.5 eV for conduction bands and -0.5 to 0 eV for valence bands) are called edge bands. Starting from a bending radius R = 10 Å, the contribution to the top of the valence band (about -0.5 to 0 eV) from the W atoms near the inner region of the ribbon starts to increase and becomes dominating at R = 6 Å. For the narrow A7WSe$_2$ nanoribbon, see SI Figure S24, the four edge W atoms dominate the DOS around the Fermi level under all the bending structures considered here, with the slightly increasing contribution from the inner three W atoms with an increase in bending.

Compared with the band structures calculated without SOC (see SI Figures S3, S15 and S22), it is obvious that the SOC effect generally induces a band splitting for those edge bands around the Fermi level and those bands close to the top of VBC. The DOS analysis shows that those edge bands are mainly contributed by the edge W atoms, and when the lower edge bands are merged into VBC, the mixed top of VBC is mainly contributed by both the edge and inner W atoms. The band structures and gaps of the nanoribbons are controlled by the states around the Fermi level from the edge and inner W atoms, the related SOC effects and bending induced strain effects across the nanoribbon. For A13WSe$_2$, the site resolved SOC energy on each W atom is shown in Figure 1q, while the complex strains on each atomic layer of the nanoribbon are shown in SI Figure S5. As can be seen, the four edge W atoms (atom indexes 1, 7, 8, and 13) have a high SOC energy among all the W atoms, and their SOC energies do not change very much (~0.005 eV) with bending. Since those edge W atoms mainly contribute to the upper edge bands for all bending radii, and to the lower edge bands before merging into VBC (R > 10 Å), the band splitting effect is resulted in for those bands. When the lower edge bands merged into VBC (R ≤ 9 Å), the inner W atoms' contribution to the top of the mixed VBC is increased, and at the same time, the SOC energies of those inner W atoms also increase. For example, the middle W atom (atom index 4) increases its SOC energy at R = 6 Å by ~5%, compared with the flat case. The increased SOC energies among the inner W atoms result in a large band splitting around the top of VBC at R = 6 Å, see Figure 1n. For the strains in the A13WSe$_2$ nanoribbon, the position 1 (12) in SI Figure S5 represents the leftmost (rightmost) pair of W atoms 1 and 8 (7 and 13), and similarly, the two outmost Se atom pairs. As can be seen, the strains in the z direction are much smaller than those in the xy plane and may play a less important role for the SOC energies. The strains in the xy plane for those outmost pairs of atoms are relatively less influenced (or varied) by bending than those inner region pairs of atoms. This can explain that the SOC energies of the four edge W atoms are relatively unchanging with bending, while the SOC energies of the inner W atoms are generally increasing with bending. Around the critical bending radius R of about 9 or 10 Å, where the lower edge bands merging into VBC, the strains in the xy plane also show a transition trend. For example, for the W atom layer, the xy strains of the inner region W pairs come back to the flat case level, when R is about 9 or 10 Å, and begin to increase largely at R < 9 Å. For the upper Se layer, the xy strains of inner pairs begin to increase strongly with bending from R < 10 Å. As for the lower Se layer, the compressive xy strains of inner atom pairs almost reach maximum from R < 10 Å. The complex strain patterns in the nanoribbons change the relative position among atoms, bond lengths and angles, induce an asymmetric internal electric field across the bent layers, and may also increase the hybridization between the d-orbital of W and the p-orbitals of both W and Se atoms in the middle region of ribbon, and hence increase the SOC energy.

## 2. Spin-polarization anisotropy with bending

The reciprocal space k point and band resolved spin polarization and magnetic moments in the x axis direction (the ribbon width direction) for the A13WSe$_2$ nanoribbon under different bending radii are plotted in Figure 2, those along axis y (perpendicular to the plane of the flat nanoribbon) are in Figure 3, and those along axis z (along the periodic direction of the nanoribbon) are in SI Figure S6. In this paper, the spin polarization is referred to the state of the spin-up (or spin-down) and the magnitude of the spin moment, not the ratio of the difference of the spin-up and spin-down to the sum of the spin-up and spin-down, since the spin moments are resolved on each k point and band, the ratio about spin-up and spin-down is not convenient. All results are for an assumed magnetic field oriented to the positive y direction, which is selected by the spin quantization axis along that direction. The overall trend is that the band spin polarization in the ribbon width direction is generally increasing with bending, and in the direction perpendicular to the plane of the flat nanoribbon, it is generally decreasing with bending, except for few k points, especially at Γ and Z (the time reversal invariant point), while in the periodic direction of the nanoribbon, it remains almost spin-unpolarized, except for few k points. The detailed analysis of the *d*-orbital decomposed spin-polarization sheds some light on the overall trend, as shown in note 2 in the SI. The spin polarization anisotropy has a subtle changing feature with bending. As can be seen, for A13WSe$_2$, there are two critical bending radii; one is $R_1 = 15$ Å and the another is $R_2 = 8$ Å or 9 Å. From flat to $R_1 = 15$ Å, the spin polarization along the width direction shows a symmetric feature, which means that the energy is degenerate, and odd and even number indexed bands have almost the same magnetic moments and opposite spin polarization. If we dope the nanoribbons with electrons and holes, doping will result in a zero net magnetic moment in the width direction.

From R = 14 Å to R = 10 Å, a spin-polarization asymmetry (or anisotropy) occurs along the ribbon width direction. As can be seen, the top of the even indexed lower edge bands has a strong spin-up polarization, while the top of the odd indexed lower edge bands has slightly lower energies and is spin-down polarized. Hole doping will result in a net magnetic moment towards the positive x direction, since a hole's spin direction is opposite to that of the electron, which is removed from the hole spot. The bottom of the upper edge bands shows a spin-up dominant feature, and especially, for $R = 11$ Å and $R = 10$ Å, the bottom of both odd and even indexed upper edge bands shows spin-up polarizations. With electron doping, the spin polarization will result in a net magnetic moment towards the negative x direction. At $R = 8$ Å, about the second critical bending radius $R_2$, the electron doping to the ribbon will cause a net spin-up magnetization along the ribbon width direction, since the bottom of the upper edge bands is spin-up dominant. However, the vicinity of the top of the valence has a relatively low spin polarization around Γ, and a slightly hole doping will result in no net magnetization. A larger hole doping will result in a net magnetization towards the positive x. At $R = 8$ Å or 9 Å, or the second critical bending radius $R_2$, the lower edge bands merge into VBC. Beyond $R_2$, for $R = 7$ Å and $R = 6$ Å, the electron doping to the ribbon still results in a net magnetization directing to the negative x, and the hole doping will also result in a net magnetization towards the negative x, since the top of the valence band is spin-down dominated under those bending radii.

The spin polarization in the y axis is generally decreasing with bending, except for some bands at Γ and Z where they show a nearly unchanged magnitude of the magnetic moment. However, the spin polarization in the y axis shows a symmetric feature among the odd and even indexed bands, resulting in no net magnetization in the y axis direction with electron and hole doping. The spin polarization in the z axis is generally small and shows a symmetric feature among the odd and even indexed bands, thus no net magnetization will occur in the ribbon's length direction with electron or hole doping. Overall, for the A13WSe$_2$ nanoribbon, the electron doped nanoribbon has a magnetization towards the negative x direction when the first critical bending radius is reached, while the hole doped nanoribbon will flip the direction of the magnetization along the ribbon width, when the bending induced merging of the edge bands into the valence band continuum occurs.

A12WSe$_2$ also shows two critical bending radii $R_1 = 12$ Å and $R_2 = 9$ Å. For $R > R_1$, the nanoribbons show a symmetric spin polarization in the nearly degenerate bands around the Fermi level. Around the second critical $R_2$, the lower edge bands merge into the valence band continuum. For both bending regions $R_2 < R < R_1$ and $R < R_2$, the upper edge bands show a spin polarization anisotropy, leading to a net magnetization along the ribbon width direction with an electron doping to the ribbon and for A12WSe$_2$ this net magnetization will flip its direction when crossing the critical $R_2$. The narrow A7WSe$_2$ nanoribbon also shows a spin polarization anisotropy in the top of the valence band at large bending. The detailed discussion about A12WSe$_2$ and A7WSe$_2$ are in note 1 in the SI.

To better understand the changing trend of the spin-polarization with bending, the d-orbital decomposition of the spin-polarization for A13WSe2 is conducted (detailed discussion in note 2 in the SI). For bent nanoribbons at $R = 13$Å and $R = 6$Å, the decomposition along the y axis has a striking difference, compared to the flat case. The two atoms in the atom pairs between the two edges produce the opposite spin direction for a given band state, leading to a totally small spin-polarization along the y axis for the bent ribbons. The edge bands in R13 (stands for $R = 13$ Å, and so on for other R's) have $d_{z^2}$, $d_{x^2-y^2}$, $d_{xz}$ and $d_{xy}$ characters, and in R6 they have $d_{z^2}$, $d_{x^2-y^2}$, $d_{xy}$, and $d_{yz}$ characters, while in the flat case, they have $d_{z^2}$, $d_{x^2-y^2}$, and $d_{xz}$ characters. This may relate to the tilting and deformation of the trigonal prisms near the ribbon edges with bending. In R6, the W atom located at the center of the ribbon contributes the majority of the spin-polarization of the top of the valence bands along the x axis, due to its location in a highly strained trigonal prism. The observation that the W atoms located on the same edge side produce the spin-polarization additively in the band states, is obeyed by all bending cases, underscoring the mirror symmetry of the nanoribbon about the central plane parallel to the yz plane. The flat and low bending nanoribbons have an additional mirror symmetry about the central W layer. This may relate to the asymmetric distribution in the magnitude of the decomposed spin-polarization, from the two atoms in the atom pairs between the two edges, in the odd number indexed and even number indexed bands.

## 3. Electron and hole doping effects with bending

To estimate the doping induced magnetization on nanoribbons, we imagine changing the Fermi level in the spin polarization resolved band structures in Figure 2 and call this method as CF (changed Fermi level), where the band structures are calculated without any doping and the nanoribbons are charge neutral. For simulating the electron doping, the changed Fermi level is raised above zero (the pristine Fermi level is set to be zero in Figure 2), and the net magnetization is calculated as the sum of the magnetic moments of the spots in the upper edge conduction bands with energies less or equal to the changed Fermi level, since those spots are populated by doping electrons under the changed Fermi level. The induced magnetization for the hole doping is also calculated in a similar manner. To evaluate the doping effect from the ab initio calculation, we also computed the band structures of A13WSe$_2$ under the three bending conditions by setting the total number of electrons in the supercell bigger (or less) than that of the neutral system and call this method as VD (VASP doping). The calculated magnetization of the A13WSe$_2$ nanoribbon under three different bending curvatures, flat, R = 13 Å, and R = 6 Å, with the CF and VD methods, is shown in Table 1, where the results are for the electron (hole) doping level of 0.12 electrons (0.12 holes) per supercell, which is equivalent to an estimated carrier doping level of ~ $1 \times 10^{20}$ cm$^{-3}$. The magnetization varying with bending from the CF and VD calculations shares a similar trend. For example, along the x direction, both CF and VD show no net magnetization for electron and hole doping for the flat ribbon. At R = 13 Å, passed the first critical bending radius $R_1$ = 15 Å, both CF and VD show a similar strength of magnetization towards the negative x with the electron doping, while a similar strength of magnetization towards the positive x with the hole doping. At R = 6 Å, passing the second critical bending radius, and the lower edge bands are merging into VBC, both CF and VD show the magnetization towards the negative x for both the electron and hole doping, confirming the direction flip of the magnetization of the hole-doped nanoribbon at large bending, while the direction of the magnetization of the electron-doped nanoribbon remains unchanged. Along the z direction, both CF and VD show no obvious magnetization for all the three bending conditions, due to the mostly negligibly small and symmetric spin polarization among bands in this direction. Along the y direction, CF gives no obvious magnetization for the three bending for either electron or hole doing and so does VD for the hole doping. Along the y direction, VD gives a non-negligible magnetization with the electron doping for flat, while at R = 13 Å and R = 6 Å VD gives no magnetization with either electron or hole doping as CF does.

The VD band structure of the flat A13WSe$_2$ nanoribbon under electron doping from the VD calculation is shown in Figure 4. It shows that the electron doping induced a significant band splitting in the upper edge bands, compared with those bands in the pristine case (Figure 1e). Since the doping electrons entering the bottom of the lower edge bands are mainly entering the d-orbital of the edge W atoms, which have a relatively high SOC energy, this induces the significant band splitting in the upper edge bands and shifts the two spin-up bands downwards in energy, resulting a spin-up dominancy at the bottom of the upper edge bands, and hence a net magnetization towards the positive y direction with the electron doping. This band spin-splitting effect is not shown in the pristine band structure (Figures 1e and 2a, and 3a), since those relevant bands are not occupied. However, this band splitting effect is not obvious in the bending radii R = 13 Å. It may be due to that the bending induces a generally decreasing in the spin polarization along the y direction and an increasing in the x direction, and the contribution from the edge W atoms to the DOS of the upper edge bands slightly decreases, while that from the inner W atoms slightly increases, see the DOS analysis of R13 in SI Figure S4. For R13, the SOC energies of all the W atoms do not change obviously, compared with those of the flat case, and the inner W atoms have much lower SOC energies than that of the edge W atoms. This results in an effective decrease in the SOC effect in the upper edge bands of the electron doped nanoribbon at R13 compared with flat, hence no obvious spin splitting in the relevant bands. For R6, the situation for the upper edge bands changes slightly compared with R13. The contribution of the inner W atoms increases slightly to those bands and the SOC energies of the inner W atoms increase significantly. This results in a slight increase in the spin splitting in the upper edge bands compared with R13. In addition, the spin polarization in the x direction becomes significant at R6, and those in the y and z directions are mostly negligible or symmetric among the odd and even indexed bands. With electron doping, a net magnetization along the negative x direction, but a negligible magnetization in the y or z direction from both CF and VD. The lower edge bands are completely merged into VBC at R6, both the contribution of the inner W atoms to the DOS of the top of the valence bands and the SOC energies of the inner W atoms increase significantly. This results in a significant band splitting and a spin polarization anisotropy in the top of the valence bands, see the spin-down dominancy in the x direction at the top of the valence band in both Figure 2k and SI Figure S7j and the resulted net magnetization towards the negative x with the hole doping in Table 1. Again, due to the mostly negligible and symmetric feature in the spin polarization in the y and z directions at R6, a hole doping results in no net magnetization in directions y and z.

## 4. Optical properties and exciton states

The change of the spin polarization with bending in the nanoribbons motivates us to calculate their optical properties. The calculated optical absorption spectra with electron hole (e-h) interaction for A7WSe$_2$ under different bending conditions are shown in Figure 5. Those without the e-h interaction are shown in SI Figure S30. In all bending cases, the absorption peaks with e-h interaction are moved to the low energy direction by about 1.0-1.5 eV, compared with those without e-h interaction, indicating the strong excitonic effect, which is a result of the reduced screening effect due to the reduced dimensionality of the nanoribbons. From the GW band structures (Figure 6 and SI Figure S29), the flat nanoribbon has a direct gap at Γ, the bent structures R16, R10, and R7 have a slightly indirect gap, and R5 has an almost direct gap at a k point (denoted as X, see Figure 6g) about 1/5 of the distance of $\Gamma - Z$ and close to the point Z. The fundamental gaps of the A7WSe$_2$ nanoribbon under different bending conditions are determined as 1.79, 1.86, 1.90, 1.75, and 1.59 eV, for the flat nanoribbon, for R16, R10, R7 and R5, respectively. Those values are more than 1.0 eV higher than those from DFT PBE, indicating the strong effect of self-energy correction of the many-body GW method. R16, R10 and R7 have relatively low absorption peaks due to their indirect gap feature. The flat nanoribbon and R5 are direct gaps and show higher absorption peaks, especially, R5 shows a much stronger absorption within the range of about 0.5-1.0 eV. The absorption edges in the optical absorption spectra of R7 and R5 are about 0.4 eV, lower than those (about 0.6 eV) of flat, R16 and R10, corresponding to the lower fundamental gaps of R7 and R5, compared to those of flat, R16 and R10.

In all the bending cases, there are several dark exciton states with lower energies than the lowest bright exciton, see SI Figure S30. To understand the spin configuration of the exciton states, we plot the spin polarization resolved GW band structures for the A7WSe$_2$ nanoribbon in Figure 6 (see SI Figure S29 for the complete plots). As can be seen, it has negligible spin polarizations along the z direction (or the ribbon length direction), consistent with the results of the DFT calculations. With an increase in bending, the spin polarization along the x direction is generally increasing, especially for the upper edge bands, while along the y direction, the spin polarization generally shows a decreasing trend, especially again for upper edge bands. Those features are also consistent with those from the DFT calculations. The overall trend for the upper edge band is that with increasing bending, the spin polarization of the upper edge bands is changing from mainly the y-direction oriented to mainly the x-direction oriented, while the spin polarization of the top of the valence bands largely remains in the x direction. The tendency that the electron and hole bands tend to orientate their spin polarizations along the same x direction and the direct gap feature at R5 can favor the formation of bright excitons and the enhanced oscillator strength, leading to a strong optical absorption.

The typical exciton states with their real spatial and reciprocal k space wavefunction distributions, decomposition of the transition bands involved, spin-polarization resolved band decomposition of their valence-to-conduction transition are shown in Figures 7 and 8. For the flat case, the band gap is direct at $\Gamma$. The exciton with the lowest energy at 0.43 eV is a dark exciton. The dark exciton is mainly due to the transition around $\Gamma$ from V1 to C2 and V2 to C1, as seen in Figure 7e. Since C1 and C2 are spin-down and spin-up around $\Gamma$, respectively, as it can be seen in Figure 6c. V1and V2 are also spin-down and spin-up, respectively, as shown in Figure 6c, this shows that this dark exciton is formed by the pair of hole and electron with like spin, resulting in a non-zero spin of the dark exciton. Note a hole has a spin opposite to that of the electron, which occupies the hole's spot in the valence band. This dark exciton spatially distributes across the width of the nanoribbon and shows a non-Frenkel exciton feature. The bright exciton at 0.69 eV has the highest oscillator strength in Peak A, mainly due to transitions around $\Gamma$ from V1 to C1 and V2 to C2, which are corresponding to the unlike spin of hole and electron, as shown in Figure 8. Due to the SOC effect, this bright exciton also has some minor contributions of the transition from the like spin pair of hole and electron, such as V1 to C2 and V1 to C3, etc. This bright exciton also shows a non-Frenkel feature and spatially distributes across the width of the nanoribbon and locates more on the two edges, since the upper edge bands C1-C4 are mainly contributed by the edge W atoms. The bright exciton at 0.9 eV in Peak B has more contributions from k points in the vicinity of $\Gamma$, and more contributions from transitions from lower V3 and V4 to higher C3 and C4. This bright exciton is also non-Frenkel like and has a mixed spin configuration in the hole and electron composition, as shown in SI Figure S31.

At $R = 10$ Å, the band gap becomes slightly indirect between the k points $\Gamma$ and Z. The exciton with the lowest energy at 0.45 eV is also dark and has mainly and almost equal contributions around $\Gamma$ from V1 to C1, V2 to C1, V1 to C2 and V2 to C2, as shown in SI Figure S32. This means that this dark exciton has almost equal contributions from pairs of hole and electron with both like spin and unlike spin. This feature is shared among the dark excitons with the lowest energy for $R = 16$ Å and $R = 7$ Å. This may be due to the slightly indirect band gap between the k points $\Gamma$ and Z, which is also a shared feature in the band structures of the cases of $R = 16$ Å, $R = 10$ Å and $R = 7$ Å. The dark excitons with a mixed spin configuration within their hole and electron pairs are potentially useful in quantum information processes. The bright exciton at 0.67 eV with the highest oscillator strength in Peak A is also mainly from transitions around $\Gamma$ from V1/V2 to C1/C2, as shown in SI Figure S33. This bright exciton also has a mixed spin configuration in the hole and electron pairs and is of non-Frenkel like. The bright exciton at 0.93 eV with the highest oscillator strength in Peak B is due to the transitions from V1-V4 to C1-C4 in a slightly spread range in k space around point Z, as shown in SI Figure S34. It has a mixed spin configuration and its wavefunction in real space shows a nodal feature.

At $R = 5$ Å, the band gap is direct at about the k point X, between $\Gamma$ and Z and closer to Z. Along the x direction, the valence bands V1-V4 at the close vicinity of X show very small spin polarization, while V2 has a slightly higher spin-down polarization than the spin-up one of V1 in the close vicinity of X, towards the $\Gamma$ direction. The exciton with the lowest energy at 0.34 eV is also a dark exciton and is mainly due to the transitions of V1 to C1 and V1/V2 to C2 in the close vicinity of X, where the hole bands have a much-reduced spin polarization, while the conduction bands have pronounced ones. This means that in this dark exciton, the hole has a much lower spin polarization than its counterpart electron. This dark exciton spatially crosses the width of the nanoribbon and extends more to the middle region of the ribbon, as shown in SI Figure S35, since at R = 5 Å, the contribution of the W atoms in the middle region to the conduction bands C1-C4 is increased. The bright exciton with the highest oscillator strength at 0.52 eV in Peak A is mainly due to transitions from V1/V2 to C1/C2 around X, as shown in SI Figure S36, especially from V2 to C2 at a k point close to X, where V2 has a relatively large spin-down polarization, compared to those k points on the side towards the Z point. This gives, in this bright exciton, a large contribution of the pair of hole and electron with unlike spin. Since the spin polarizations of the electron and hole are all aligned along the x direction (those along the y and z directions are negligibly small), together with the direct band gap around X, this may explain the enhanced absorption at the energy of this exciton. Like the dark exciton, this bright exciton also shows an extended spatial distribution, due to the contribution of the middle region W atoms to the conduction bands. The bright exciton with the highest oscillator strength at 0.70 eV in Peak B has a nodal feature in k space. It has a more contribution from transitions V1/V2 to C1/C2 from k points away from X, such as the middle region between Γ and Z, and around Γ. Also, it has a mixed spin configuration and spatially extended, as shown in SI Figure S37.

By bending the $A7WSe_2$ nanoribbons, the induced non-uniform strains change the SOC effect, band structures, and spin polarization of bands dramatically. Bending results in a large tunability in the exciton states and optical absorption in the energy range of about 0.4 to 1.5 eV, which is within the technologically relevant near infrared region. The rich spin configuration in the dark and bright excitons in the bent nanoribbons can also have potential in the application of the spin-based quantum information processes.

**Conclusion**

In summary, from first principles, we calculated the band structures, band gaps, and projected spin-polarization of the semiconducting armchair $WSe_2$ nanoribbons under different bending curvatures and electron/hole doping with density functional approximations (DFA), and their optical absorption and excitonic states with the many-body perturbation GW and BSE methods. These calculations show that both the upper edge bands (the conduction bands just above the Fermi level) and the lower edge bands (the valence bands just below the Fermi level) are mainly derived from the *d*-orbital of the edge W atoms. With increasing bending curvature, the contributions from the W atoms in the middle region of the ribbon to those edge bands increase. We find that the band gap of the nanoribbon can be changed from direct to indirect, or indirect to direct under appropriate bending. With a magnetic field aligned perpendicular to the ribbon plane, the spin polarization perpendicular to the ribbon plane is generally decreasing with bending, the one along the ribbon width direction is generally increasing with bending, while the one along the ribbon length direction is generally small. In the undoped DFA calculation, for the relatively wider nanoribbons $A12WSe_2$ and $A13WSe_2$, there are two critical bending radii $R_1$ and $R_2$. For a bending curvature radius $R > R_1$, the nanoribbons show a symmetric spin polarization in the nearly degenerate bands around the Fermi level. Around the second critical $R_2$, the lower edge bands merge into the valence band continuum. For both bending regions $R_2 < R < R_1$ and $R < R_2$, the upper edge bands show a spin polarization anisotropy (or asymmetry), leading to a net magnetization along the ribbon width direction with an electron doping to the ribbon and for $A12WSe_2$ this net magnetization will flip its direction when crossing the critical $R_2$. The top of the valence bands can also show the spin polarization anisotropy with bending, resulting in a net magnetization with a hole doping to the system and for $A13WSe_2$ this net magnetization will flip its direction when crossing the critical $R_2$. The doped DFA calculation basically shows the similar characteristics for A13WSe$_2$. It also shows that for the electron doped flat A13WSe$_2$, it has a net magnetization perpendicular to the ribbon plane, due to the electron occupation of the upper edge bands and the strong SOC effect in the edge bands. The narrow A7WSe$_2$ nanoribbon also shows a spin polarization anisotropy in the top of the valence band at large bending. We also calculated armchair WS$_2$ and WTe$_2$ nanoribbons under different bending curvatures (results not shown) and they all show the similar behaviors of spin-polarization anisotropy within the bands around the Fermi level. It suggests that the doped and bent tungsten dichalcogenide nanoribbons can be used in compact and controllable magnetic nanodevices and spintronics. The optical absorption of A7WSe$_2$ presents a large tunability with bending within the technologically important near infrared region of about 0.4 to 1.5 eV, showing an enhanced absorption at a large bending, that may be due to the enhanced spin alignment between the underlying electrons and holes and the direct gap feature at the large bending condition. The bright excitons in A7WSe$_2$ generally show a mixed spin configuration in the electron and hole pairs, while the lowest energy dark excitons show a various spin configuration with bending, potential for possible applications in spin-based quantum information processes.

**Methods**

Density functional theory (DFT) calculations were conducted in the Vienna Ab initio Software Package (VASP)[28] with projector augmented-wave pseudopotentials.[29, 30] The Perdew-Burke-Ernzerhof (PBE)[31] functional with SOC was used to calculate the band structures of nanoribbons. The vacuum layer of more than 12 Å is added along the direction of nanoribbon width and inserted along the direction perpendicular to the 2D surface of the nanoribbon, to avoid the interactions between the nanoribbon and its periodic images. The energy cutoff is 580 eV. The k-point mesh of $1 \times 1 \times 24$ was used for all nanoribbons. The nanoribbons were built from the pre-relaxed WSe$_2$ monolayer, which was relaxed with PBE and has the in-plane lattice constants $a = b = 3.321$ Å. All nanoribbons were fully structurally relaxed with PBE with all forces less than 0.01 eV/Å. During the relaxation, the x and y coordinates of the two outermost metal atoms on the two edge sides were fixed, while their coordinates along the ribbon axis direction, which is the z direction, and all the coordinates of other atoms were allowed to relax. The supercell vectors a and b along the x and y axes were also allowed to relax. The definitions and the calculating formula of strains in the nanoribbons can be found in the reference 23. The doping calculations were conducted in VASP with the selected total number of electrons in the supercell bigger (or less) than that of a neutral one and the monopole correction was included. The G$_0$W$_0$ [32, 33] and G$_0$W$_0$+BSE[34] calculations were conducted in BerkeleyGW[32] by pairing with Quantum ESPRESSO.[35] The wavefunction energy cutoff is 70 Ry (~950 eV). The energy cutoff for the epsilon matrix is 18 Ry (~240 eV). The k-point mesh of $1 \times 1 \times 36$ and both valence and conduction bands of 6 was set for optical absorption calculations. The band number for summation is 1100. The correction of the exact static remainder and the wire Coulomb truncation for 1D systems were also used.

**Supporting Information**

Technical Note 1 for spin polarization anisotropy analyses for A12WSe2 and A7WSe2 nanoribbons, note 2 for the d-orbital decomposition analyses for A13WSe2, Table S1 for the five observations to the band structure map of the d-orbital decomposition of the spin-polarization along the x and y axes for the A13WSe2 nanoribbon under different bending conditions, Figures S1 and S2 for the relaxed structures of the A12WSe2 and A7WSe2 nanoribbons, Figures S3-S7, and S15-S28 for the band structures with SOC and/or without SOC, density of states (DOS), site resolved SOC energies, strain analyses, band-structure mapped spin-polarizations, for A13WSe2, A12WSe2 and A7WSe2, Figures S8-S14 for the d-orbital decomposition analyses for A13WSe2, and Figures S29-S37 for GW band structures and spin polarization, optical absorption and exciton spectra, the exciton analyses at the different energies for the A7WSe2 nanoribbons.

**Author Contributions**

H.T. and A.R. designed the research; H.T. and S.N. conducted the computation and analyzed the data; J.M.B. did some exciton analyses; H.T. wrote the manuscript with discussions with A.R., S.N., and L.Y.; all authors discussed the results and contributed to the manuscript.

**Notes**

The authors declare no competing financial interest. The data that support the findings of this study are available from the corresponding author upon reasonable request.

**Acknowledgement**

This material is based upon work supported by the U.S. Department of Energy, Office of Science, Office of Basic Energy Sciences, under Award Number DE-SC0021263. This research used resources of the National Energy Research Scientific Computing Center, a DOE Office of Science User Facility supported by the Office of Science of the U.S. Department of Energy under Contract No. DE-AC02-05CH11231.

Table 1. The calculated magnetization (in unit Bohr magneton $\mu_B$ per supercell) along the x, y, and z directions for the A13WSe$_2$ nanoribbon under different bending from methods CF (changed Fermi level) and VD (vasp doping). The electron (hole) doping level is 0.12 electrons (0.12 holes) per supercell. Flat represents a flat nanoribbon, and R13 for bending radius R = 13 Å and so on.

| axis | flat | | | | R13 | | | | R6 | | | |
|---|---|---|---|---|---|---|---|---|---|---|---|---|
| | e-doping | | h-doping | | e-doping | | h-doping | | e-doping | | h-doping | |
| | CF | VD | CF | VD | CF | VD | CF | VD | CF | VD | CF | VD |
| x | 0.000 | 0.000 | 0.000 | 0.000 | -0.010 | -0.008 | +0.019 | +0.014 | -0.050 | -0.018 | -0.038 | -0.022 |
| y | 0.000 | +0.094 | 0.000 | 0.000 | 0.000 | 0.000 | 0.000 | 0.000 | +0.001 | 0.000 | 0.000 | 0.000 |
| z | 0.000 | 0.000 | 0.000 | 0.000 | 0.000 | 0.000 | 0.000 | 0.000 | -0.001 | 0.000 | 0.000 | 0.000 |

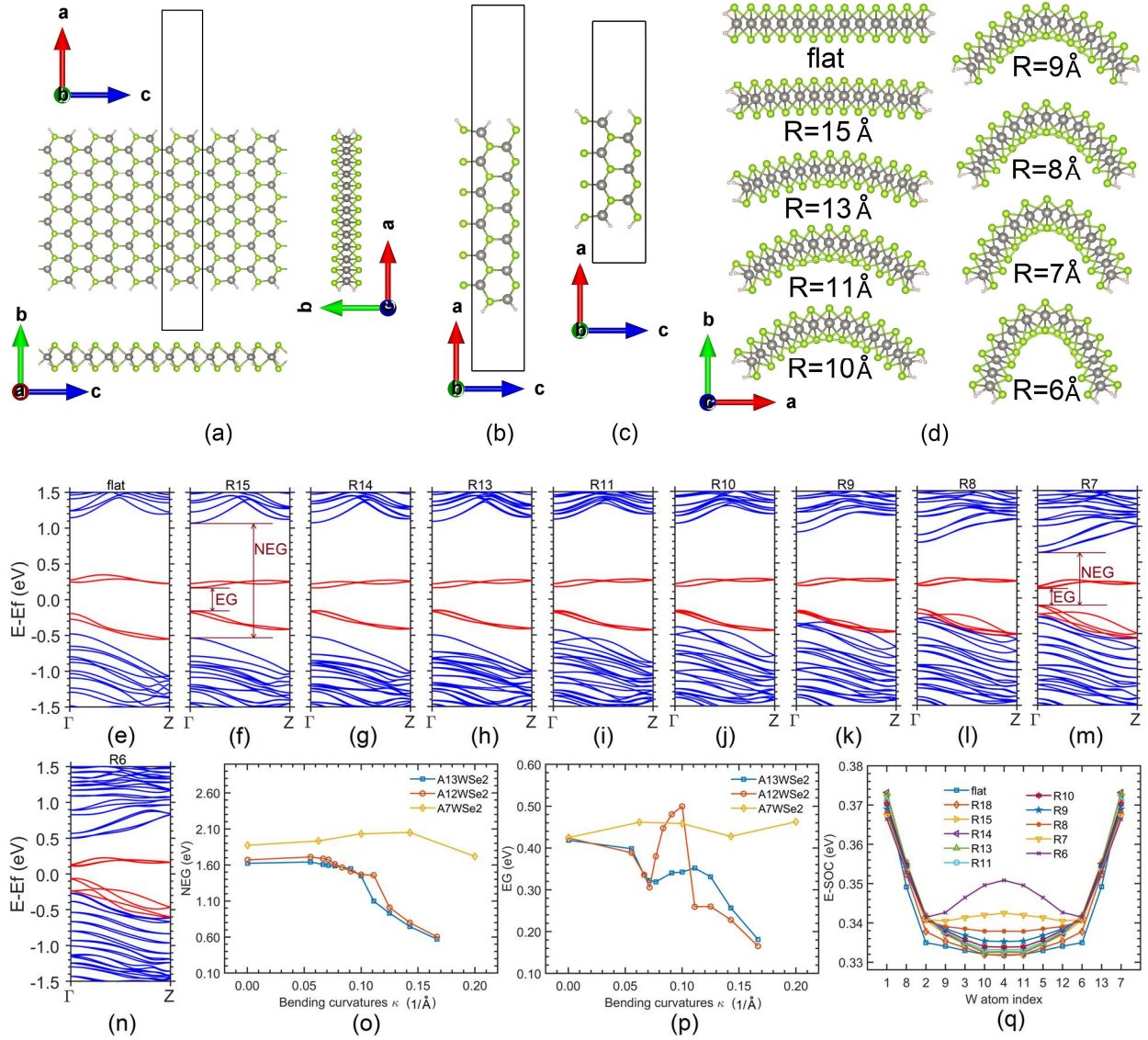

Figure 1. The structures of the armchair WSe$_2$ nanoribbons, their relaxed structures, band structures, band gaps, and the SOC energies on metal sites under different bending curvature radii R. The flat nanoribbons A$n$WSe$_2$ with $n$=13, 12, and 7 are shown in (a), (b) and (c), respectively. $n$ in A$n$WSe$_2$ represents the number of W atoms in the periodic supercell, whose unit vectors a, b, and c are aligned with axes x, y, and z, respectively. Blue balls represent W atoms, green balls are for Se atoms and small white balls are H atoms. The side views of the A13WSe$_2$ nanoribbon are also shown in (a). Panel (d) shows the relaxed structures of the A13WSe$_2$ nanoribbon under flat, R = 15 Å, R = 13 Å, R = 11 Å, R = 10 Å, R = 9 Å, R = 8 Å, R = 7 Å, and R = 6 Å. Panels (e), (f), (g), (h), (i), (j), (k), (l), (m), and (n) are the calculated band structures with SOC of the A13WSe$_2$ nanoribbon for flat ribbon, R = 15 Å, R = 14 Å, R = 13 Å, R = 11 Å, R = 10 Å, R = 9 Å, R = 8 Å, R = 7 Å, and R = 6 Å, respectively. The edge bands around the Fermi level are shown in red and other bands are in blue. The edge band gap (EG) is defined as the difference between the conduction band minimum (CBM) and the valence band maximum (VBM), while the non-edge band gap (NEG) is defined as the one between the bottom of the conduction band

continuum (CBC) and the top of the valence band continuum (VBC), as shown in panel (f). With large bending (R ≤ 9 Å), the lower edge bands just below the Fermi level merge into VBC and in these cases, EG is the difference between CBM and VBC, as shown in panel (m). The NEG and EG of the three nanoribbons A$n$WSe$_2$ with $n$=13, 12, and 7 as the function of bending curvature $\kappa$ ($\kappa = 1/R$) are in (o) and (p), respectively. Panel (q) shows the SOC energy on each W atom in the A13WSe$_2$ nanoribbon as a function of bending. The energy is negative, and the magnitude of the energy is plotted. The W atoms are placed across the ribbon width. The atom indexes within the middle region of panel (q) represent the W atoms located in the middle region of the nanoribbon, while the two on the leftmost (index 1) and the rightmost (index 7) represent the two edge W atoms in the nanoribbon.

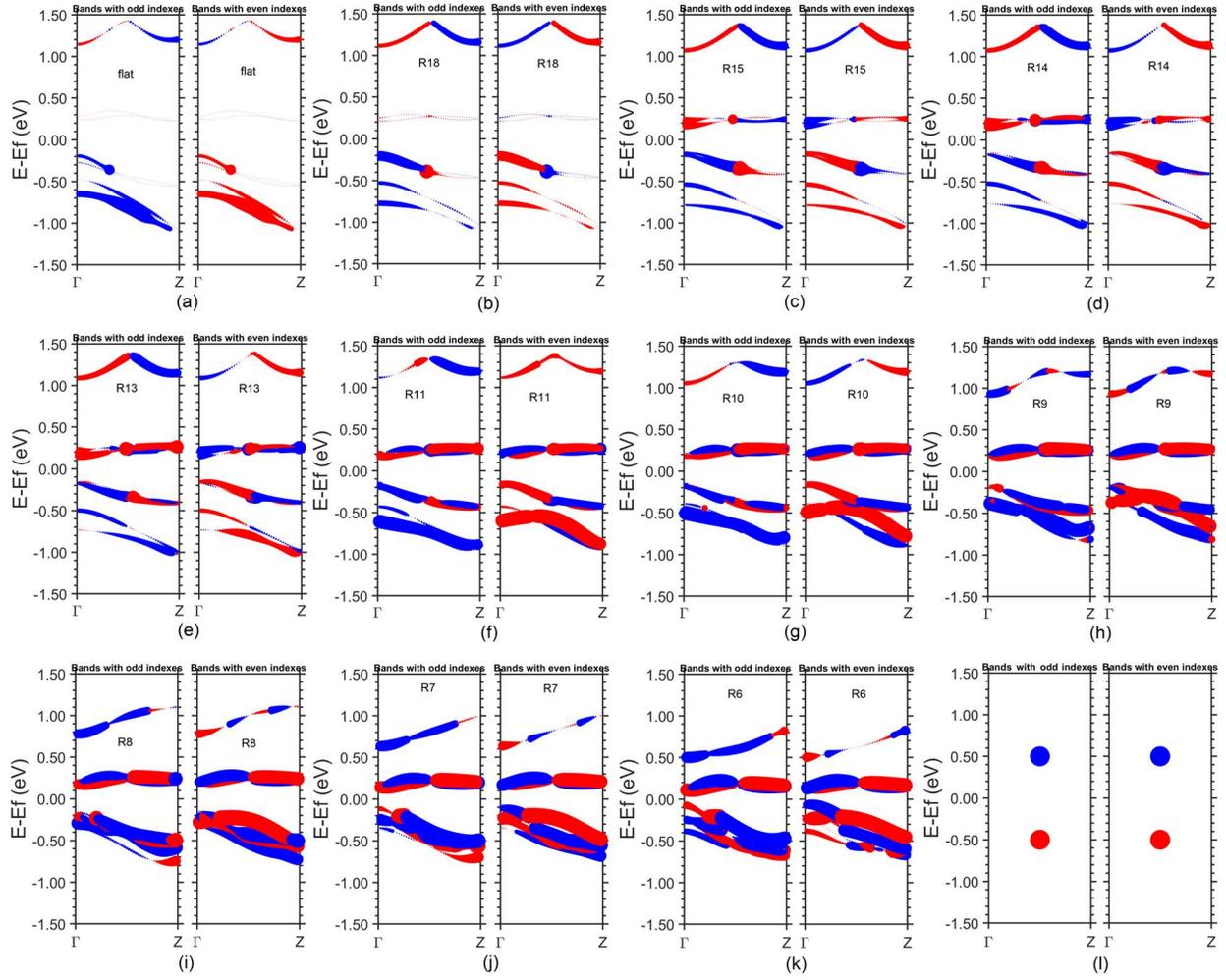

Figure 2. The k point and band resolved spin polarization and magnetic moments along the width direction (x axis) of the A13WSe$_2$ nanoribbon under different bending curvature radii R. Panels (a), (b), (c), (d), (e), (f), (g), (h), (i), (j), (k), and (l) are for flat, R = 18 Å, R = 15 Å, R = 14 Å, R = 13 Å, R = 11 Å, R = 10 Å, R = 9 Å, R = 8 Å, R = 7 Å, and R = 6 Å, respectively. The bending radius is also marked in each panel, i.e., R18 means R = 18 Å and so on for others. In each panel, the left subplot is for odd number indexed bands, while the right one is for the even number indexed bands. The red (blue) color represents the spin-up (spin-down) spin polarization, with the spin-up (spin-down) pointing to the negative (positive) direction of the x axis. The magnitude of the magnetic moment (in the unit of Bohr magneton $\mu_B$) at each k point and each band is proportional to the size of the spot. Panel (l) represents the reference size of the spots with the magnetic moment of 1 $\mu_B$ (red) or -1 $\mu_B$ (blue). Note that all the spin-up or spin-down directions calculated in the VASP code are in the so-called internal coordinate system, which is not identical to the xyz coordinate system of the supercell and they are related by a transformation matrix. However, the x, y, and z in all figures presented in this paper are referred to those in the xyz coordinate system of the supercell. This explains why a spin-up along the x (or z) axis is not pointing to the positive x (or z) direction, but for y, it is.

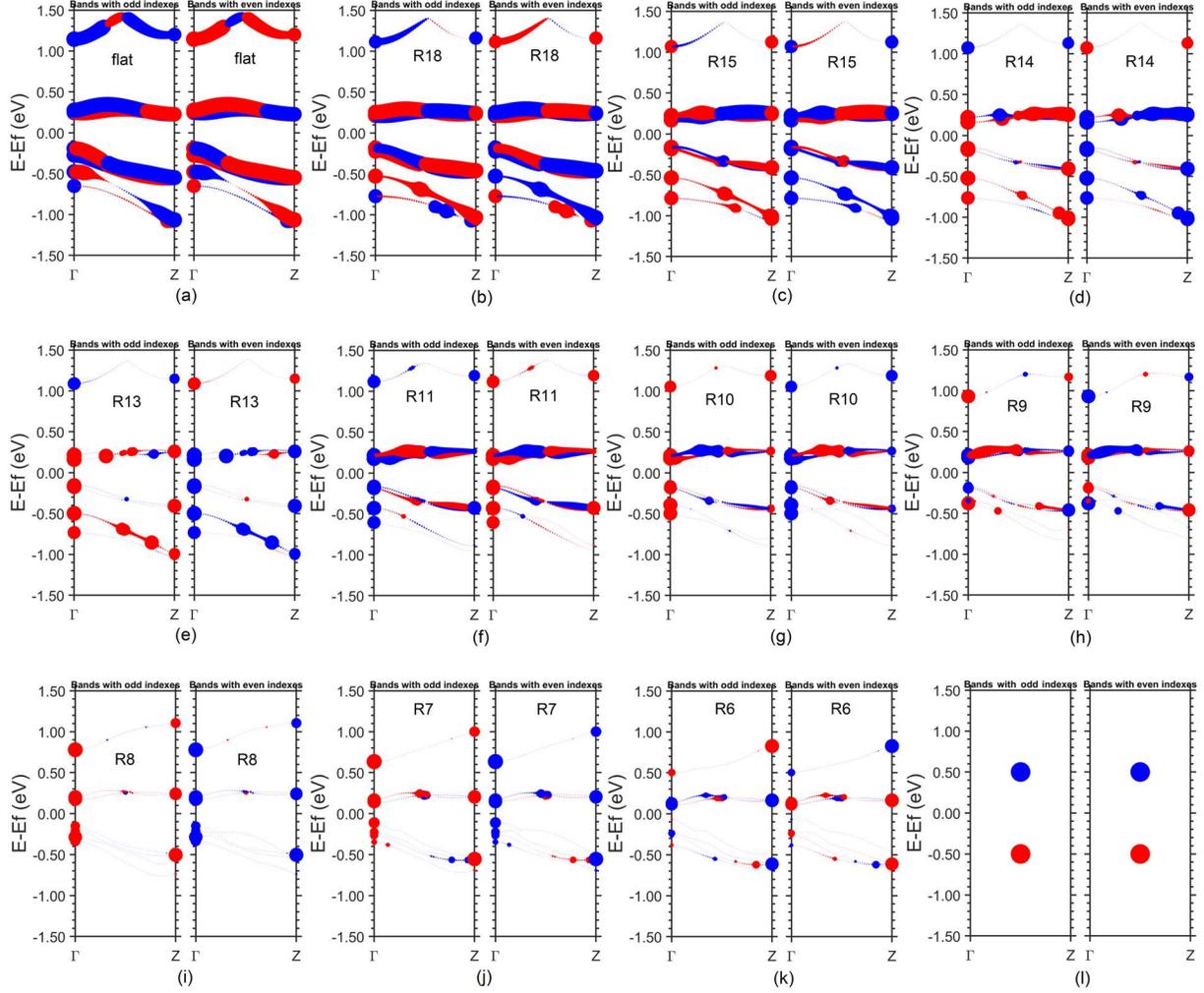

Figure 3. The k point and band resolved spin polarization and magnetic moments along the y axis (perpendicular to the plane of the flat nanoribbon) of the A13WSe$_2$ nanoribbon under different bending curvature radii R. Panels (a), (b), (c), (d), (e), (f), (g), (h), (i), (j), (k), and (l) are for flat, R = 18 Å, R = 15 Å, R = 14 Å, R = 13 Å, R = 11 Å, R = 10 Å, R = 9 Å, R = 8 Å, R = 7 Å, and R = 6 Å, respectively. In each panel, the left subplot is for odd number indexed bands, while the right one is for the even number indexed bands. The red (blue) color represents the spin-up (spin-down) spin polarization, with the spin-up (spin-down) points to the positive (negative) direction of the y axis. The magnitude of the magnetic moment (in the unit of Bohr magneton $\mu_B$) at each k point and each band is proportional to the size of the spot. Panel (l) represents the reference size of the spots with the magnetic moment of 1 $\mu_B$ (red) or -1 $\mu_B$ (blue).

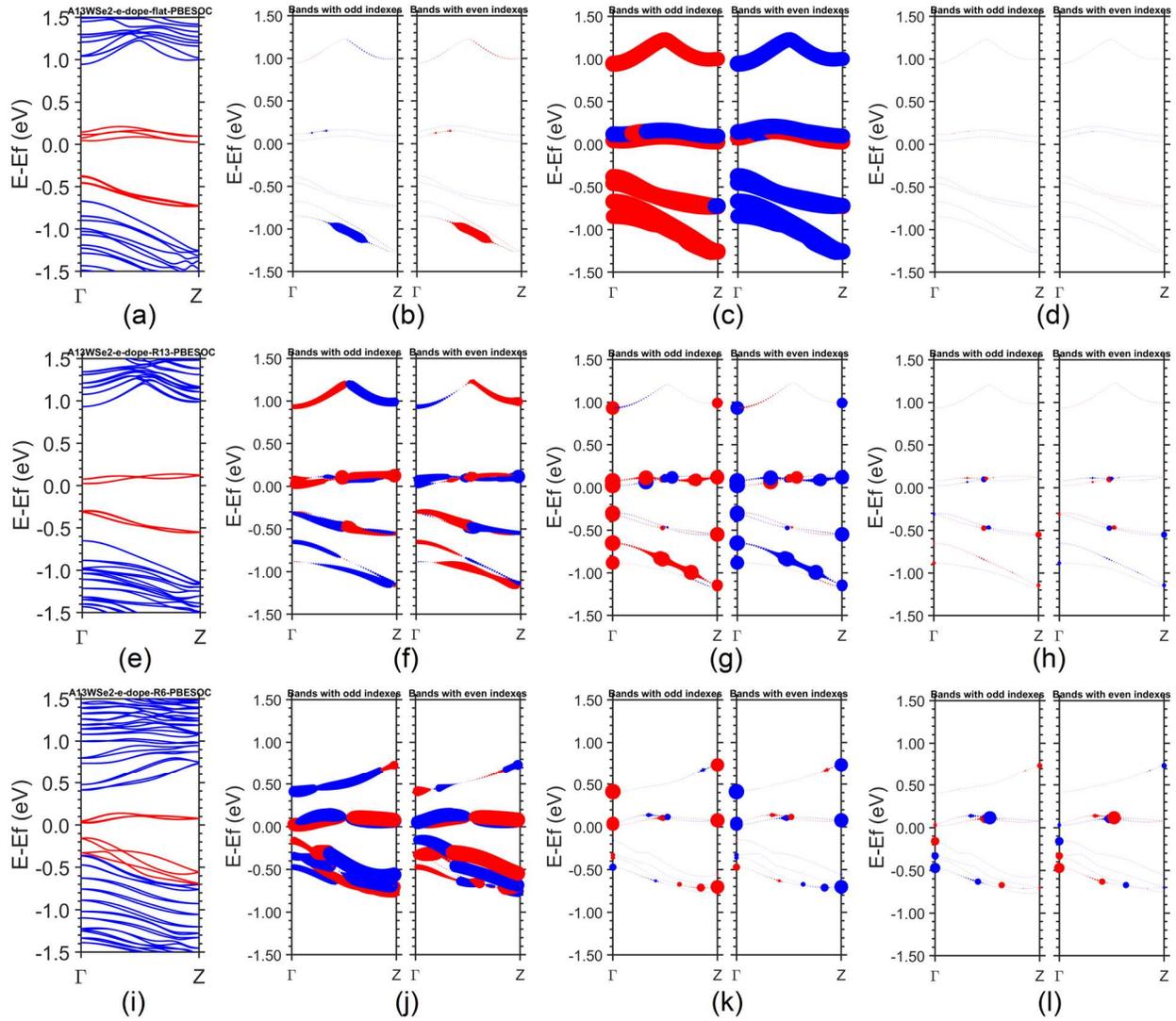

Figure 4. The band structures and k point and band resolved spin polarizations of the A13WSe$_2$ nanoribbon under the electron doping of 0.12 electrons per supercell and under three bending conditions. Panels (a), (b), (c), and (d) in the top row are the band structure, spin polarization along the x axis, spin polarization along the y axis, and spin polarization along axis z, respectively, for the flat nanoribbon. In panels (b)-(d), the odd number indexed bands are plotted on the left side and the even numbered ones are on the right side with the same energy scale, for clarity. The directions to which a spin-up (or a spin-down) polarization points in the two spatial axes x and y are the same as those defined in Figures 2 and 3. The direction to which a spin-up (or a spin-down) polarization points in the z axis, is the same as that defined in Figure S6. The middle (bottom) row is for bending radius R = 13 Å (R = 6 Å). The figure panels in the last two rows are arranged similarly as in the top row. The figure for the hole doping is in Figure S7 in the Supplemental Information.

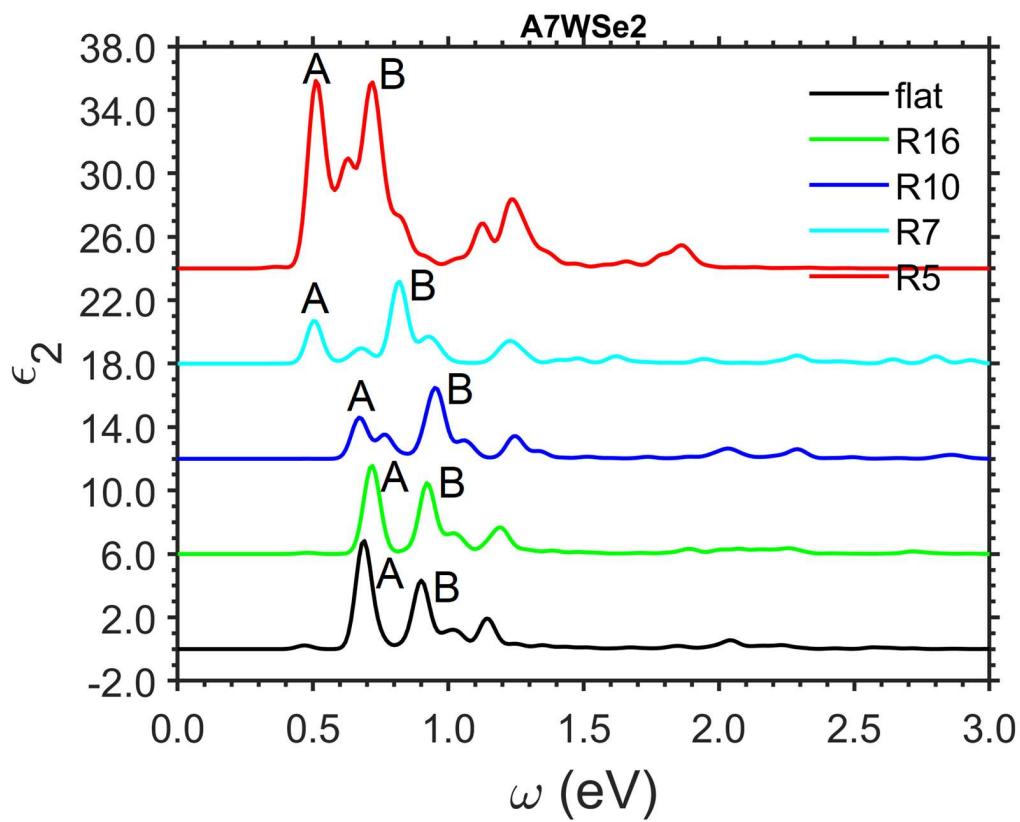

Figure 5. The optical absorption spectra of the A7WSe$_2$ nanoribbon under different bending radii calculated with GW (at the DFT of PBE) and BSE. A constant Gaussian broadening of 28 meV is used in all calculations. Only the results with the electron and hole interaction are shown. The two adjacent curves are shifted vertically by 6 eV for clarity. The results without the electron and hole interaction are in Figure S30 in the Supplemental Information.

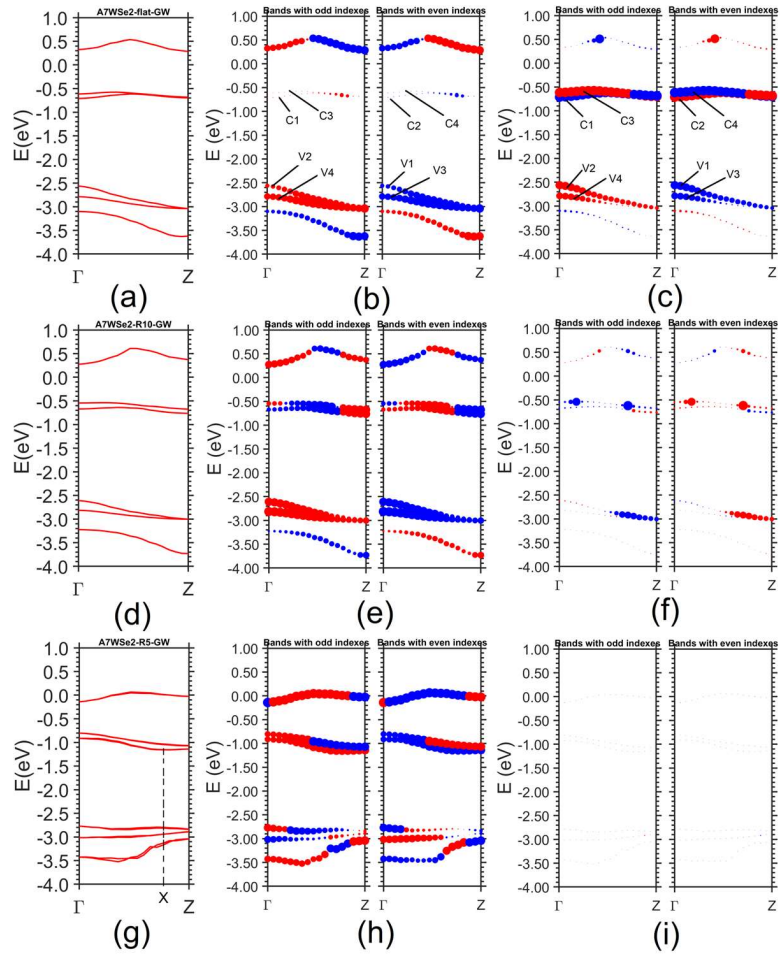

Figure 6. The $G_0W_0$ band structures and k point and band resolved spin polarizations of the $A7WSe_2$ nanoribbon under three bending conditions. Panels (a), (b), and (c) in the top row are the $G_0W_0$ band structures, spin polarization along axis x, and spin polarization along axis y, respectively, for the flat nanoribbon. In panels (b) and (c), the odd number indexed bands are plotted on the left side and the even numbered ones are on the right side with the same energy scale, for clarity. Panes (b) and (c) also show the second kind of band labels, i. e. C1, C2, … and V1, V2, …, which are for labeling the conduction bands (C1, C2, …) counted upwards from the Fermi level, and for the valence bands (V2, V2, …) counted downwards from the Fermi level. The highest valence band is labelled as V1 and the lowest conduction band is labelled as C1. The middle (bottom) row is for bending radius R = 10 Å (R = 5 Å). The figure panels in the last two rows are arranged similarly as in the top row. The directions to which a spin-up (or a spin-down) polarization points in the two spatial axes x and y are the same as those defined in Figures 2 and 3. The plots for the spin polarization along axis z and other bending radii are in Figure S29 in the Supplemental Information.

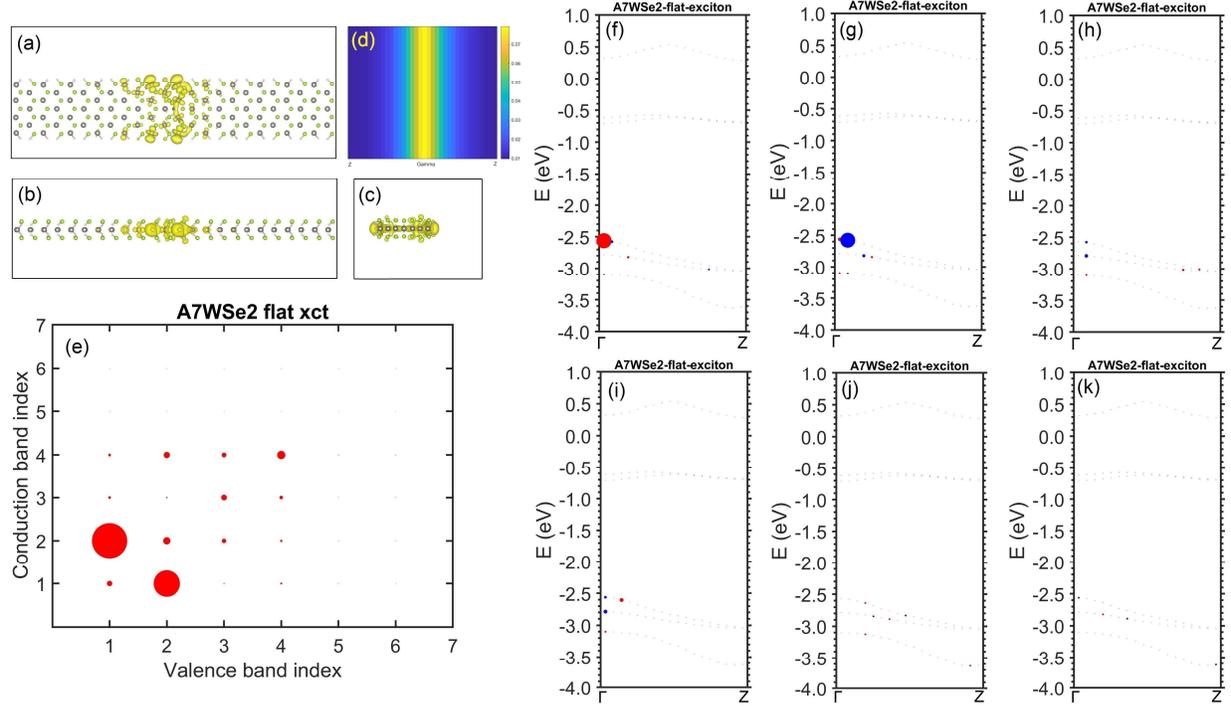

Figure 7. The dark exciton at the energy of 0.43 eV of the flat A7WSe$_2$ nanoribbon. (a)-(c) show the three views of the squared modulus of wavefunction of this exciton in real space. (d) shows the band-summed envelope function of the exciton, $\sum_{v,c}|A_{v,c}(k)|^2$, representing the distribution of excitons in k-space. (e) shows the decomposition of the transition bands involved in this exciton, and the size of the spots proportional to $\sum_k|A_{v,c}(k)|^2$ and represents the contributing weight from the v-c pair. (f), (g), (h), (i), (j), and (k) represent the contribution from all valence bands to the conduction band C1, C2, C3, C4, C5, and C6, respectively. (f)-(k) are in the same energy scale as Figure 6. The spot size in the valence states in (f)-(k) is proportional to the modulus square of the exciton envelope function $|A_{v,c}(k)|^2$, representing the contribution from that valence band and k point to the corresponding conduction band. The red (blue) color is for spin-up (spin-down) in (f)-(k).

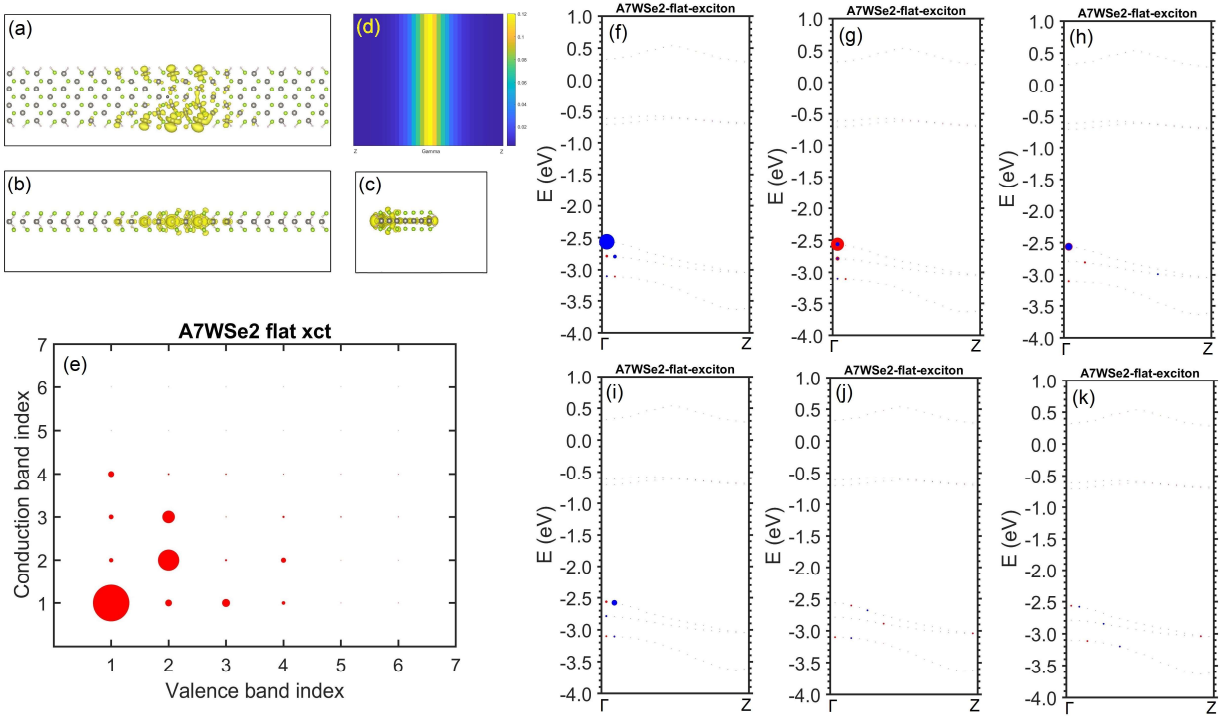

Figure 8. The bright exciton at the energy of 0.69 eV of the flat A7WSe$_2$ nanoribbon. (a)-(k) are plotted and arranged in the same way as in Figure 7.